\documentclass[abstract,headings=normal]{scrartcl}

\usepackage[automark]{scrpage2}
\usepackage[english]{babel}
\usepackage[T1]{fontenc}
\usepackage[applemac]{inputenc}
\usepackage{epic,eepic,graphicx}
\usepackage{tikz}
\usepackage{amsmath}
\usepackage{amssymb}
\usepackage{amsthm}
\usepackage{subfig}
\usepackage{accents}

\def\Black(#1,#2){\put(#1,#2){\circle*{\rrr}}}
\def\White(#1,#2){\put(#1,#2){\circle{\rrr}}}

\newcommand{\bbC}{\mathbb{C}}
\newcommand{\bbZ}{\mathbb{Z}}
\newcommand{\bbR}{\mathbb{R}}

\newcommand{\cL}{\mathcal{L}}

\newtheorem{theorem}{Theorem}

\newtheorem{definition}[theorem]{Definition}
\theoremstyle{remark}

\graphicspath{{./graphics/}}
\DeclareGraphicsExtensions{.pdf}

\begin{document}

\title{Variational symmetries \\ and pluri-Lagrangian systems}
\author{Yuri B. Suris}
\date{}
\maketitle

\vspace{-1.5cm}

\begin{center}
{
Institut f\"ur Mathematik, MA 7-2, Technische Universit\"at Berlin, \\
Str. des 17. Juni 136, 10623 Berlin, Germany\\
E-mail: {\tt suris@math.tu-berlin.de}
}
\end{center}

\begin{abstract}
We analyze the relation of the notion of a pluri-Lagrangian system, which recently emerged in the theory of integrable systems, to the classical notion of variational symmetry, due to E.~Noether.
\end{abstract}

\section{Introduction}

\begin{figure}[htbp]
\begin{center}
\includegraphics[width=0.5\textwidth]{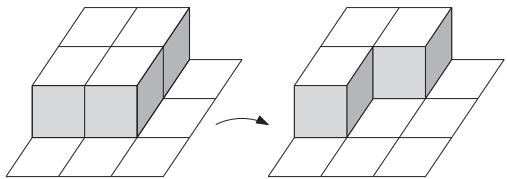}
\caption{Local move of a quad-surface involving one three-dimensional cube}
\label{Fig: local moves}
\end{center}
\end{figure}

In the last decade, a new understanding of integrability of discrete systems as their multi-dimensional consistency has been a major breakthrough \cite{BS1}, \cite{N}. This led to classification of discrete 2-dimensional integrable systems (ABS list)  \cite{ABS}, which turned out to be rather influential. According to the concept of multi-dimensional consistency, integrable two-dimensional systems can be imposed in a consistent way on all two-dimensional sublattices of a lattice $\bbZ^m$ of arbitrary dimension. This means that the resulting multi-dimensional system possesses solutions whose restrictions to any two-dimensional sublattice are generic solutions of the corresponding two-dimensional system. To put this idea differently, one can impose the two-dimensional equations on any quad-surface in $\bbZ^m$ (i.e., a surface composed of elementary squares), and transfer solutions from one such surface to another one, if they are related by a sequence of local moves, each one involving one three-dimensional cube, like the moves shown of Fig. \ref{Fig: local moves}.

A further fundamental conceptual development was initiated in \cite{LN1} and deals with variational (Lagrangian) formulation of discrete multi-dimensionally consistent systems. Solutions of any ABS equation on any quad surface $\Sigma$ in $\bbZ^m$ are critical points of a certain action functional $S_\Sigma=\int_\Sigma\cL$ obtained by integration of a suitable discrete Lagrangian 2-form $\cL$. It was observed in \cite{LN1} that the critical value of the action remains invariant under local changes of the underlying quad-surface, or, in other words, that the 2-form $\cL$ is closed on solutions of quad-equations, and it was suggested to consider this as a defining feature of integrability. Results of \cite{LN1}, found on the case-by-case basis for some equations of the ABS list, have been extended to the whole list and were given a more conceptual proof in \cite{BS2}. Subsequently, this research was pushed in various directions: for multi-field two-dimensional systems \cite{LN2, ALN}, for dKP, the fundamental three-dimensional discrete integrable system \cite{LNQ}, and for the discrete time Calogero-Moser system, an important one-dimensional integrable system \cite{YLN}. A general theory of multi-time one-dimensional Lagrangian systems has been developed in \cite{S}. In \cite{BPS} the general structure of multi-time Euler-Lagrange equations for two-dimensional Lagrangian problems was studied, and it was shown that the corresponding 2-form $\cL$ is closed not only on solutions of (non-variational) quad-equations, but also on general solutions of the corresponding Euler-Lagrange equations.

As argued in the latter paper, the original idea of \cite{LN1} has significant precursors. These include:
\begin{itemize}
\item {\em Theory of pluriharmonic functions} and, more generally, of pluriharmonic maps \cite{R, OV, BFPP}. By definition, a pluriharmonic function of several complex variables $f:\bbC^m\to\bbR$ minimizes the Dirichlet functional $E_\Gamma=\int_\Gamma |(f\circ \Gamma)_z|^2dz\wedge d\bar z$ along any holomorphic curve in its domain $\Gamma:\bbC\to\bbC^m$. Differential equations governing pluriharmonic functions (and maps) are heavily overdetermined. Therefore it is not surprising that they belong to the theory of integrable systems.
\item {\em Baxter's Z-invariance} of solvable models of statistical mechanics  \cite{Bax1, Bax2}. This concept is based on invariance of the partition functions of solvable models under elementary local transformations of the underlying planar graphs. It is well known (see, e.g., \cite{BoMeSu}) that one can identify planar graphs underlying these models with quad-surfaces in $\bbZ^m$. On the other hand, the classical mechanical analogue of the partition function is the action functional. This suggests the relation of Z-invariance to the concept of closedness of the Lagrangian 2-form, at least at the heuristic level. This relation has been made mathematically precise for a number of models, through the quasiclassical limit \cite{BMS1, BMS2}.
\item The classical notion of {\em variational symmetry},  going back to the seminal work of E.~Noether \cite{Noether}, turns out to be directly related to the idea of the closedness of the Lagrangian form in the multi-time.
\end{itemize}

The first of these precursors motivates the following term which was proposed in \cite{BPS} to describe the general framework of the deveopment we are speaking about: {\em pluri-Lagrangian systems}. A $d$-dimensional pluri-Lagrangian problem can be set as follows: given a $d$-form $\cL$ on an $m$-dimensional space (called multi-time, $m>d$), whose coefficients depend on a sought-after function $x$ of $m$ independent variables (called field), find those fields $x$ which deliver critical points to the action functionals $S_\Sigma=\int_\Sigma\cL$ for {\em any} $d$-dimensional manifold $\Sigma$ in the multi-time.

The intention of the present note is to explain the relation of this notion to the third of the above mentioned precursors, namely to the notion of variational symmetries. For this aim, we recall the necessary definitions in section \ref{sect: var}, illustrating them with one of the most familiar examples, the sine-Gordon equation and its variational symmetry given by the modified KdV equation. Then, in section \ref{sect: closed} we establish the relation of these classical notions with the idea of closedness of the Lagrangian form. Finally, in section \ref{sect: pluri} we present the Euler-Lagrange equations for two-dimensional pluri-Lagrangian problems of second order, and establish the pluri-Lagrangian structure of the sine-Gordon equation.

\section{Variational symmetries}
\label{sect: var}

Let us start with reminding the notion of a variational symmetry of Lagrangian differential equations, introduced in the seminal paper by E.~Noether \cite{Noether} (see also a detailed historical discussion in \cite{Kosmann}).

We consider the differential algebra of functions $u^\alpha$ ($\alpha=1,\ldots,q$) of independent variables $x^i$ ($i=1,\ldots,p$). It has generators $u^\alpha_I$, $I=(i_1,\ldots,i_n)$ being a multiindex. The derivation $D_j$, understood as a full derivative w.r.t. $x^j$, acts on generators according to $D_ju^\alpha_I=u^\alpha_{I+e_j}$. Thus, for any differential function $f$ we have
\[
 D_jf=\frac{\partial f}{\partial x^j}+\sum u^\alpha_{I+e_j}\ \frac{\partial f}{\partial u^\alpha_I}.
\]

We now define more general derivations (generalized vector fields). In what follows, we only consider ``vertical'' (or evolutionary) generalized vector fields, i.e., those acting on dependent variables only. This is done, on one hand, for simplicity of notation, and, on the other hand, because in the discrete case only these are relevant, due to the absence of changes of independent variable.

\begin{definition}{\bf (Generalized evolutionary vector field)} A generalized evolutionary vector field generated by the set of $q$ differential functions $\varphi[u]=(\varphi^1[u],\ldots,\varphi^q[u])$ is given by
\begin{equation}\label{eq: gen vector field}
D_\varphi=\sum_{\alpha,I} \varphi^\alpha_I\ \frac{\partial}{\partial u^\alpha_I},\quad \varphi^\alpha_I=D_I\varphi^\alpha=
D_1^{i_1}\ldots D_n^{i_n}\varphi^\alpha.
\end{equation}
\end{definition}
Usually \cite[eq. (5.6)]{Olver}, this is called an (infinite) prolongation of the generalized evolutionary vector field
\[
\sum_\alpha \varphi^\alpha \frac{\partial}{\partial u^\alpha}.
\]
The following is an adaptation of Definition 5.51 from \cite{Olver}.
\begin{definition}{\bf (Variational symmetry)}  A generalized evolutionary vector field (\ref{eq: gen vector field}) is called a {\em variational symmetry} of an action functional
\[
S[u]=\int L[u]dx^1\ldots dx^p,
\]
if the action of $D_\varphi$ on the Lagrange function $L$ is a complete divergence, that is, if there exist functions $M_1[u],\ldots,M_p[u]$ such that
\begin{equation}\label{eq: var sym}
D_\varphi L={\rm Div}\ M=\sum_{i=1}^p D_iM_i.
\end{equation}
\end{definition}
The intention of this definition is clear: the integral of a complete divergence (by fixed boundary values) vanishes, so $D_\varphi$ preserves the value of the action functional. The following statement (see \cite[Theorem 5.53]{Olver}) justifies the previous definition.
\begin{theorem} If a generalized evolutionary vector field $D_\varphi$ is a variational symmetry of the action functional $S$, then it is a generalized symmetry of the Euler-Lagrange equations $\delta L/\delta u=0$.
\end{theorem}

{\bf Example.} Sine-Gordon equation, $p=2$, $q=1$:
\begin{equation}\label{eq: SG}
u_{xy}=\sin u,
\end{equation}
is the Euler-Lagrange equation for
\begin{equation}\label{eq: SG L}
L[u]=\frac{1}{2}u_xu_y-\cos u.
\end{equation}
We show that the (prolongation of the) evolutionary vector field $\varphi\, \partial/\partial u$ with
\begin{equation}\label{eq: phi}
\varphi[u]=u_{xxx}+\frac{1}{2}u_x^3
\end{equation}
is a variational symmetry for the sine-Gordon equation. The corresponding computation is mentioned in \cite[p. 336]{Olver}, but is not presented there in detail, being replaced by a less direct method. We show that 
\begin{equation}\label{eq: SG var sym}
    D_\varphi L=D_xN+D_yM
\end{equation}
with the following differential functions:
\begin{eqnarray}
M[u] & = & \dfrac{1}{2}\varphi u_x-\dfrac{1}{8}u_x^4+\dfrac{1}{2}u_{xx}^2, \label{eq: SG M}\\
N[u] & = & \dfrac{1}{2}\varphi u_y -\dfrac{1}{2}u_x^2\cos u - u_{xx}(u_{xy}-\sin u). \label{eq: SG N}
\end{eqnarray}

Indeed, we compute:
\begin{equation}\label{eq: SG var sym aux1}
D_\varphi L=\frac{1}{2}(\varphi_yu_x+\varphi_xu_y)+\varphi\sin u,
\end{equation}
and
\begin{eqnarray}
D_y M+D_x N & = & \dfrac{1}{2}\varphi_yu_x+\dfrac{1}{2}\varphi u_{xy}-\dfrac{1}{2}u_x^3u_{xy}+u_{xx}u_{xxy}
        \nonumber\\
        &   & +\dfrac{1}{2}\varphi_xu_y+\dfrac{1}{2}\varphi u_{xy}-u_xu_{xx}\cos u +\dfrac{1}{2}u_x^3\sin u \nonumber\\
        &   & -u_{xxx}(u_{xy}-\sin u)-u_{xx}(u_{xxy}-u_x\cos u) \nonumber\\
        & = & \dfrac{1}{2}(\varphi_yu_x+\varphi_xu_y)+\varphi \sin u
              +\Big(\varphi-\dfrac{1}{2}u_x^3-u_{xxx}\Big)(u_{xy}-\sin u).  \qquad\quad \label{eq: SG var sym aux2}
\end{eqnarray}
Comparing (\ref{eq: SG var sym aux1}) and (\ref{eq: SG var sym aux2}), we see that identity (\ref{eq: SG var sym}) is satisfied under the choice of the differential function $\varphi$ as in (\ref{eq: phi}). By the Noether's theorem, existence of a variational symmetry leads to the corresponding conservation law for the sine-Gordon equation:
\begin{eqnarray*}
\varphi\;\frac{\delta L}{\delta u} & = & \varphi\left(\frac{\partial L}{\partial u}-D_x\frac{\partial L}{\partial u_x}-D_y\frac{\partial L}{\partial u_y}\right)\\
& = & \varphi\,\frac{\partial L}{\partial u}+\varphi_x\,\frac{\partial L}{\partial u_x}+\varphi_y\,\frac{\partial L}{\partial u_y}-D_x\left(\varphi\frac{\partial L}{\partial u_x}\right)-D_y\left(\varphi\frac{\partial L}{\partial u_y}\right)\\
& = & D_\varphi L-D_x\left(\varphi\frac{\partial L}{\partial u_x}\right)-D_y\left(\varphi\frac{\partial L}{\partial u_y}\right)\\
& = & D_x\left(N-\varphi\frac{\partial L}{\partial u_x}\right)+D_y\left(M-\varphi\frac{\partial L}{\partial u_y}\right)\\
& = & D_x\left(N-\frac{1}{2}\varphi u_y\right)+D_y\left(M-\frac{1}{2}\varphi u_x\right)\\
& = & -D_x\left(\dfrac{1}{2}u_x^2\cos u + u_{xx}(u_{xy}-\sin u)\right)+D_y\left(-\dfrac{1}{8}u_x^4+\dfrac{1}{2}u_{xx}^2\right),
\end{eqnarray*}
which can be also found in \cite[p. 336]{Olver}.

\section{Variational symmetries and closedness of multi-time Lagrangian forms}
\label{sect: closed}

Now, we would like to promote an alternative point of view. In the standard approach, reproduced in the previous section, equation (\ref{eq: var sym}) is a certain (differential-)algebraic property of the vector field $D_\varphi$. However, this way of thinking about this equation ignores one of the main interpretations of the notion of ``symmetry'', namely the interpretation as a commuting flow. In this interpretation, one introduces a {\em new independent variable} $z$ corresponding to the ``flow'' of the generalized vector field $D_\varphi$,
\begin{equation}\label{eq: flow sym}
D_z u^\alpha=\varphi^\alpha[u],
\end{equation}
and considers {\em simultaneous solutions} of the Euler-Lagrange equations $\delta L/\delta u=0$ and of the flow (\ref{eq: flow sym}) as functions of $p+1$ independent variables $x^1,\ldots,x^p,z$. Then equation (\ref{eq: var sym}) reads
\begin{equation}\label{eq: var sym new}
D_z L-\sum_{i=1}^p D_iM_i=0.
\end{equation}
The key observation is that equation (\ref{eq: var sym new}) is nothing but the {\em closedness condition} of the following $p$-form in the $(p+1)$-dimensional space:
\begin{equation}
\cL=L[u]\ dx^1\wedge \ldots\wedge dx^p-\sum_{i=1}^p (-1)^i M_i[u]\ dz\wedge dx^1\wedge\ldots\wedge \widehat{dx^i}\wedge \ldots \wedge dx^p.
\end{equation}
Thus, we are led to define the {\em extended action functional}
\begin{equation}\label{eq: extended action}
S_\Sigma=\int_\Sigma \cL,
\end{equation}
where $\Sigma$ is some $p$-dimensional surface (with boundary) in the $(p+1)$-dimensional space of independent variables $x^1,\ldots,x^p,z$. In particular, the action $S_\Sigma$ over the hypersurface $\Sigma\subset\{z={\rm const}\}$ is the original action $S$. Equation (\ref{eq: var sym new}) means that the extended action does not depend on local changes of the $p$-dimensional integration surface $\Sigma$ preserving boundary. Of course, this statement only holds {\em on simultaneous solutions} of the Euler-Lagrange equations $\delta L/\delta u=0$ and of the flow (\ref{eq: flow sym}).
\smallskip

{\bf Example.} To clearly see this in our above example of the sine-Gordon equation, we re-write the previous computations in our new notation, i.e., we replace $\varphi$ by $u_z$. We have:
\begin{equation}\label{eq: SG ext action}
\cL =  L[u]\ dx\wedge dy-M[u]\ dz\wedge dx-N[u]\ dy\wedge dz,
\end{equation}
where
\begin{eqnarray}
L[u] & = & \dfrac{1}{2}u_xu_y-\cos u, \label{eq: SG Lagr L}\\
M[u] & = & \dfrac{1}{2}u_xu_z-\dfrac{1}{8}u_x^4+\dfrac{1}{2}u_{xx}^2, \label{eq: SG Lagr M}\\
N[u] & = & \dfrac{1}{2}u_yu_z-\dfrac{1}{2}u_x^2\cos u-u_{xx}(u_{xy}-\sin u). \label{eq: SG Lagr N}
\end{eqnarray}
Then the previous computation tells us that
\begin{equation}\label{eq: closure}
L_z-(M_y+N_x)=-\Big(u_z-\dfrac{1}{2}u_x^3-u_{xxx}\Big)(u_{xy}-\sin u).
\end{equation}
Thus, the form $\cL$ is closed as soon as $u_z=\frac{1}{2}u_x^3+u_{xxx}$. This shows us once again that the modified KdV equation
\begin{equation}\label{eq: mKdV}
    u_z=u_{xxx}+\frac{1}{2}u_x^3
\end{equation}
is a variational symmetry of the sine-Gordon equation (\ref{eq: SG}).

The remarkable factorized form of the right-hand side of (\ref{eq: closure}) shows that it also vanishes as soon as $u_{xy}=\sin u$. This suggests that the above relation could be reversed, namely, that the sine-Gordon equation should be a variational symmetry of the modified KdV equation, as well. Two facts apparently stand in the way of this interpretation: first, modified KdV equation is not Lagrangian, and, second, sine-Gordon equation is not evolutionary. Nevertheless, this interpretation is still possible. To show this, we first observe that the function $M[u]$ from (\ref{eq: SG Lagr M}) can be considered as a Lagrangian for the action $S_\Sigma$ over the hypersurface $\Sigma\subset\{y={\rm const}\}$. The corresponding Euler-Lagrange equation $\delta M/\delta u=0$ is
\begin{equation}\label{eq: mKdV diff}
u_{zx}-\frac{3}{2}u_x^2 u_{xx}-u_{xxxx}=0,
\end{equation}
the differentiated form of modified KdV. It is this equation for which we want to declare the derivation $D_y$ as a variational symmetry. To overcome the difficulty that $D_y$ is not an evolutionary vector field (i.e., that $u_y$ is not defined by our differential equations), we observe that we only need to define the action of $D_y$ on the Lagrangian $M$. However, the latter function does not contain $u$ alone, but only its derivatives (of degree 1 w.r.t $z$ and of higher degrees w.r.t. $x$). For such functions the formula
\[
D_yf=u_{yz}\frac{\partial f}{\partial u_z}+\sum_{I:\ i_1\ge 1} u_{Iy}\, \frac{\partial f}{\partial u_I}
\]
works perfectly as an evolutionary vector field. Indeed, one can use the equation
\[
u_{yz}=u_{xxxy}+\frac{3}{2}u_x^2u_{xy}=(\sin u)_{xx}+\frac{3}{2}u_x^2\sin u=u_{xx}\cos u+\frac{1}{2}u_x^2\sin u,
\]
which is obtained from (\ref{eq: mKdV}) by differentiation upon use of the sine-Gordon equation, as well as relations $u_{Iy}=D_{I-e_1}\sin u$ for multiindices $I$ with $i_1\ge 1$.

\section{Pluri-Lagrangian structure of the sine-Gordon equation}
\label{sect: pluri}
Next, we regard the multi-time Euler-Lagrange equations for the pluri-Lagrangian problem with  the Lagrangian 2-form $\cL$. We will not give the complete derivation, but restrict ourselves to the statement which covers our main example in this note, namely the sine-Gordon equation.

\begin{theorem} Consider a pluri-Lagrangian problem with the 2-form
\[
\cL=\sum_{i<j}L_{ij}[u]dx^i\wedge dx^j,
\]
where $L_{ij}=-L_{ji}$ are differential functions depending on the second jet of the sought-after function $u=u(x^1,\ldots,x^n)$, so that $L_{ij}[u]=L_{ij}(u,u_k,u_{km})$. The system of {\em multi-time Euler-Lagrange equations} consists of:
\begin{eqnarray}
\frac{\delta L_{ij}}{\delta u} & = & 0, \label{eq: pluri 1}\\
\frac{\delta L_{ij}}{\delta u_k} & = & 0, \quad k\notin \{i,j\} \label{eq: pluri 2}\\
\frac{\delta L_{ij}}{\delta u_{km}} & = & 0,\quad k,m\notin\{i,j\} \label{eq: pluri 3}\\
\frac{\delta L_{ij}}{\delta u_j} & = & p_i {\rm \;\; does \;\; not\;\; depend\;\; on\;\;} j\neq i, \label{eq: pluri 4}\\
\frac{\delta L_{ij}}{\delta u_{jk}} & = &  p_i^k {\rm \;\; does \;\; not\;\; depend\;\; on\;\;} j\neq i, \label{eq: pluri 5}
\end{eqnarray}
and
\begin{equation}\label{eq: pluri 6}
\frac{\delta L_{ij}}{\delta u_{ij}}+\frac{\delta L_{jk}}{\delta u_{jk}}+\frac{\delta L_{ki}}{\delta u_{ki}}=0.
\end{equation}
Here the following notations are used:
\begin{eqnarray}
\frac{\delta L_{ij}}{\delta u} & := & \frac{\partial L_{ij}}{\partial u}-D_i\frac{\partial L_{ij}}{\partial u_i}
-D_j\frac{\partial L_{ij}}{\partial u_j}+D_i^2\frac{\partial L_{ij}}{\partial u_{ii}}
+D_iD_j\frac{\partial L_{ij}}{\partial u_{ij}}+D_j^2\frac{\partial L_{ij}}{\partial u_{jj}}, \qquad\\
\frac{\delta L_{ij}}{\delta u_k} & := & \frac{\partial L_{ij}}{\partial u_k}-D_i\frac{\partial L_{ij}}{\partial u_{ik}}-D_j\frac{\partial L_{ij}}{\partial u_{jk}}, \\
\frac{\delta L_{ij}}{\delta u_{km}} & := & \frac{\partial L_{ij}}{\partial u_{km}}.
\end{eqnarray}
\end{theorem}
Equations (\ref{eq: pluri 1})--(\ref{eq: pluri 3}) can be derived by considering action over coordinate planes as surfaces $\Sigma$, while equations (\ref{eq: pluri 4})--(\ref{eq: pluri 6}) are derived by considering general surfaces $\Sigma$ and are less obvious.
\medskip

{\bf Example:}  sine-Gordon equation. We have:
\begin{eqnarray}
L=L_{12} & = & \dfrac{1}{2}u_xu_y-\cos u, \label{eq: SG Lagr L12}\\
M=L_{13} & = & \dfrac{1}{2}u_zu_x-\dfrac{1}{8}u_x^4+\dfrac{1}{2}u_{xx}^2, \label{eq: SG Lagr L13}\\
N=-L_{23} & = & \dfrac{1}{2}u_zu_y -\dfrac{1}{2}u_x^2\cos u-u_{xx}(u_{xy}-\sin u). \label{eq: SG Lagr L23}
\end{eqnarray}
For these Lagrangians, the above system of multi-time Euler-Lagrange equations reduces to
\begin{eqnarray}
&& u_{xy}=\sin u, \label{eq: u xy}\\
&& u_{xz}=\frac{3}{2}u_x^2 u_{xx}+u_{xxxx}, \label{eq: u xz}\\
&& u_{yz}=u_{xx}\cos u+\frac{1}{2}u_x^2\sin u, \label{eq: u yz}\\
&& u_{xxy}=u_x\cos u, \label{eq: u xxy} \\
&& u_z=\frac{1}{2}u_x^3+u_{xxx}. \label{eq: u z}
\end{eqnarray}
Indeed, for $n=3$ the above system consists of 19 equations. With the present choice of $L_{ij}$, the majority of these equations are satisfied identically. The non-trivial equations are derived from the following ones:
\begin{itemize}
\item[--] equation (\ref{eq: u xy}) is obtained from two equations, namely, from $\delta L/\delta u=0$ and from $\delta N/\delta u_{xx}=0$;
\item[--] equation (\ref{eq: u xz}) is obtained from $\delta M/\delta u=0$;
\item[--] equation (\ref{eq: u yz}) is obtained from $\delta N/\delta u=0$;
\item[--] equation (\ref{eq: u xxy}) is obtained from $\delta N/\delta u_x=0$;
\item[--] equation (\ref{eq: u z}) is obtained from
\[
\frac{\delta N}{\delta u_y}+\frac{\delta M}{\delta u_x}=
\left(\frac{\partial N}{\partial u_y}-D_y\frac{\partial N}{\partial u_{yy}}-D_z\frac{\partial N}{\partial u_{yz}}\right)
+\left(\frac{\partial M}{\partial u_x}-D_z\frac{\partial M}{\partial u_{xz}}-D_x\frac{\partial M}{\partial u_{xx}}\right)=0.
\]
\end{itemize}
It remains to notice that equations (\ref{eq: u xz})--(\ref{eq: u xxy}) are corollaries of (\ref{eq: u xy}) and (\ref{eq: u z}), derived by differentiation.
\begin{theorem}
Multi-time Euler-Lagrange equations for the pluri-Lagrangian problem with the 2-form (\ref{eq: SG ext action}) with the components (\ref{eq: SG Lagr L})--(\ref{eq: SG Lagr N}) consist of the sine-Gordon equation (\ref{eq: u xy}) and the modified KdV equation (\ref{eq: u z}). On simultaneous solutions of these equations, the 2-form $\cL$ is closed.
\end{theorem}
It is remarkable that multi-time Euler-Lagrange equations are capable of producing evolutionary equations.

\section{Conclusions}

In subsequent publications, we will address the following problems:
\begin{itemize}
\item[--] To derive multi-time Euler-Lagrange equations for pluri-Lagrangian problems of arbitrary order, i.e., for forms $\cL$ depending on jets of arbitrary order.
\item[--] To extend the classical De Donder-Weyl theory of calculus of variations to the pluri-Lagrangian context.
\item[--] To elaborate on the pluri-Lagrangian structure of classical integrable hierarchies, like the KdV or, more generally, Gelfand-Dickey hierarchies. Note that in the monograph \cite{Dickey}, which is, in my opinion, one of the best sources on the Lagrangian field theory and whose program, according to the foreword, is ``that the book is about hierarchies of integrable equations rather than about individual equations'', it is the Lagrangian part (chapters 19, 20) that only deals with individual equations. The reason for this is apparently the absence of the concept of pluri-Lagrangian systems.
\item[--] To establish a general relation of pluri-Lagrangian structure to more traditional notions of integrability.
\item[--] To study the general relation of pluri-Lagrangian structure to Z-invariance of statistical-mechanical problems, via quasi-classical limit, as exemplified in \cite{BMS1, BMS2}.
\end{itemize}

\medskip

This research is supported by the DFG Collaborative Research Center TRR 109 ``Discretization in Geometry and Dynamics''.

\bibliographystyle{amsalpha}

\end{document}